%% file: main.tex
\documentclass[aps+-,prl,twocolumn,amsfonts,amssymb,amsmath,nofootinbib,superscriptaddress]{revtex4-2}

\usepackage{graphicx} 

\usepackage[separate-uncertainty]{siunitx} 

\usepackage[T1]{fontenc} 
\usepackage[english]{babel}

\usepackage{hyperref} 
\usepackage[capitalize]{cleveref} 

\usepackage{float} 

\begin{document}

\title{Micromotion compensation using dark and bright ions}

\author{Orr Barnea}
\email{orr.barnea@weizmann.ac.il}
\author{Dror Einav}
\author{Jonas Drotleff}
\author{Idan Hochner}
\author{Ziv Meir}
\email{ziv.meir@weizmann.ac.il}
\affiliation{Department of Physics of Complex Systems, Weizmann Institute of Science, Rehovot 761001, Israel}

\begin{abstract}
Stray electric fields induce excess micromotion in ion traps, limiting experimental performance. 
We present a new micromotion-compensation technique that utilizes a dark ion in a bright-dark-bright linear ion crystal. Stray electric fields in the radial plane of the trap deform the crystal axially. We exploit the mode softening near the transition to the zig-zag configuration to increase our sensitivity dramatically.  
We corroborate our results with a modified ion-displacement compensation method using a single bright ion. Our modification allows us to compensate stray fields on the 2D radial plane from a 1D measurement of the ion position on the camera.
Both methods require only a fixed imaging camera and continuous ion-fluorescence detection.
As such, they can be readily implemented in virtually any ion-trapping experiment without additional hardware modifications.
\end{abstract}

\maketitle 

\section{Introduction} 
Excess micromotion (EMM) is present in any radio-frequency (RF) ion trap. It is the residual motion of an ion in its equilibrium position driven by the RF oscillating fields~\cite{Berkeland1998}.
While EMM could be used to one's advantage in some exceptional cases~\cite{navon2013addressing}, typically, compensating EMM below some threshold is a necessary condition for most ion-trapping experiments. A few prominent examples are experiments involving the interaction between neutral atoms and trapped ions where EMM sets the energy scale of the interaction~\cite{tomza2019cold}, optical-atomic clocks where uncompensated EMM could limit the clock's accuracy~\cite{Brewer2019Al+10-18}, and optical trapping of ions where the optical dipole force is overwhelmed by uncompensated stray electric fields in the trap~\cite{huber2014far}.  

Several techniques for detecting and compensating EMM were developed in the last few decades. Each technique has its advantages and disadvantages when compared to others~\cite{Keller2015}. For example, the resolved-sideband method~\cite{Berkeland1998, Keller2015} and its recent extension~\cite{arnold2024enhanced} only detect EMM parallel to the laser's k-vector and require the development of coherent control tools. Here, we focus on a technique for which advanced coherent control tools are unnecessary.  

The minimal requirement for trapping an ion is the continuous illumination with a laser to induce ion fluorescence and an imaging system to collect the emitted photons on a camera. This is the typical starting point of any ion-trapping experiment. With only these tools at hand, EMM compensation can be achieved by observing the position of the ion on the camera for different radio-frequency trapping strengths~\cite{Berkeland1998,schneider2012influence,gloger2015ion, saito2021measurement}. While 2D information on the ion position is readily extracted from a fluorescence image of an ion, the information regarding the position along the imaging axis is less accessible~\cite{schneider2012influence}. It requires automated mechanical scanning of the focus~\cite{gloger2015ion, saito2021measurement} or engineering of the imaging point-spread function~\cite{zhou2024tracking}, which are not standard techniques in ion imaging.  

Here, we use a dark ion ($^{44}\text{Ca}^+$) embedded in a crystal of bright ions ($^{40}\text{Ca}^+$) as a sensitive detector of radial stray electric fields and for EMM compensation. Since the radial confinement in linear RF traps depends on the mass of the ion, mixed-species ions, which carry different masses, will acquire a differential shift in their position due to radial stray electric fields~\cite{Mokhberi_2015}. This will lead to a deformation of the ion crystal detectable by fluorescence imaging. Minimizing the crystal deformation inevitably indicates the suppression of stray electric fields in the ion trap. 

In a linear crystal of bright ions embedded with a dark ion, a stray radial electric field can be detected by observing the deformation of the crystal along its axis (axial axis), which can be sensitive to a displacement of the dark ion in both radial directions. Thus, by reading the position of the bright ions along the axial axis on the 2D imaging plane, we are sensitive to radial displacements of the dark ion that are both on and out of the imaging plane. 

Axial deformation due to radial displacement is quadratic in nature. Hence, we need to enhance the sensitivity of our method further. We do so by working near the transition from a linear to a zig-zag configuration in a bright-dark-bright three-ion crystal~\cite{Kaufmann2012Precise}. Near the transition, the frequency of the radial bending mode approaches zero (mode softening) \cite{Fishman2008Structural}, significantly enhancing crystal deformation for a given stray field.

Mixed-species ion crystals are ubiquitous in novel ion-trapping quantum technologies such as mixed-species optical atomic clocks~\cite{Brewer2019Al+10-18, Hausser2025}, quantum-logic control of molecular ions~\cite{Wolf2016, Lin2020QuantumMolecule, Sinhal2020Quantum-nondemolitionMolecules, holzapfel2024quantumcontrolsinglemathrmh2} and highly charged ions~\cite{king2022optical}, and mixed-species quantum computation~\cite{Tan2015Multi-elementQubits, Ballance2015HybridIsotopes}, to name a few. Therefore, techniques for compensating EMM in these mixed-species crystals are of high importance~\cite{barrett2003}.  

\begin{figure*}
    \centering
    \includegraphics[width=\linewidth]{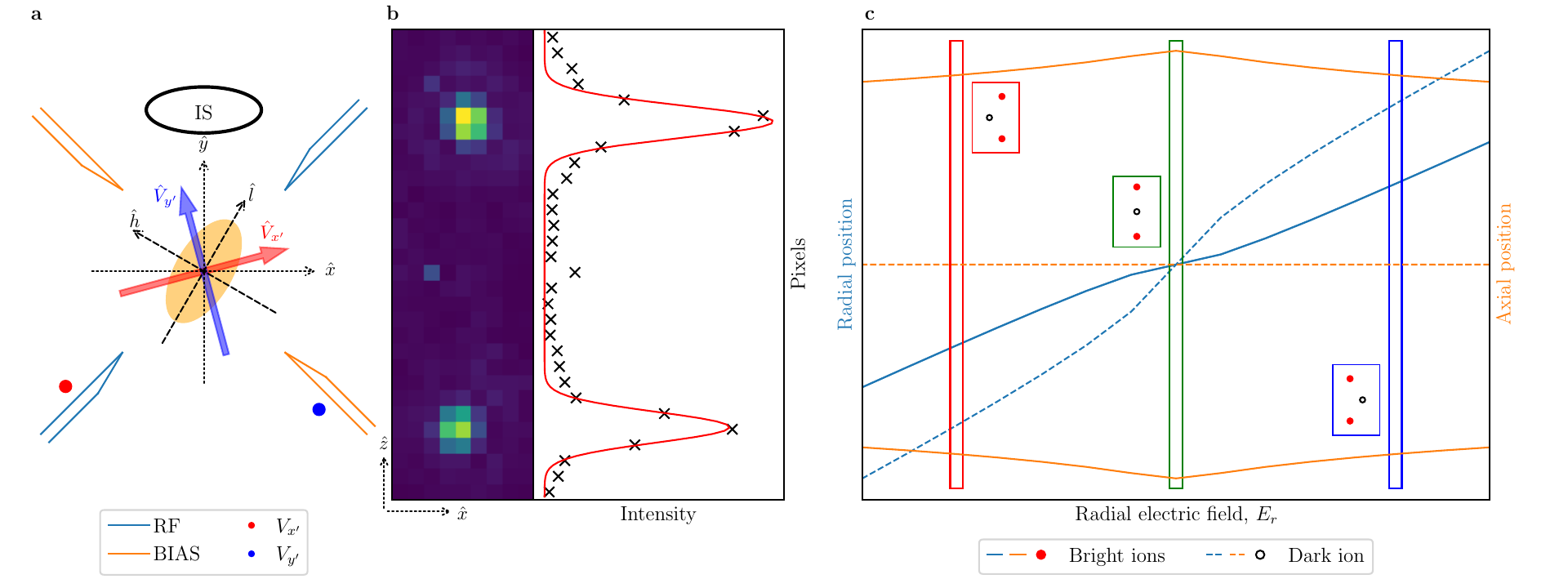}
    \caption{Deformation of a bright-dark-bright Coulomb crystal due to uncompensated EMM. a) Sketch of a radial cross section of the ion trap showing the RF (light blue) and BIAS (orange) trap blades, which control the RF and bias trapping-fields amplitudes. Two compensation rods, $V_{x'}$ (red) and $V_{y'}$ (blue), create a radial electric field along the $\hat{V}_{x'}$ and $\hat{V}_{y'}$ directions. The imaging system (IS) captures the position of the ions along one projection of the radial direction ($\hat{x}$) and the axial direction ($\hat{z})$, as shown in (b). The displacement of the ions due to external field is dependent on the orientation (dashed axes) and ratio (orange ellipse) of the high ($\hat{h}$) and low ($\hat{l}$) radial modes. b) Sample image of the ions captured by the IS (1 sec exposure time, 5 images average). The position of the ions is found by fitting the intensity to a Gaussian in both the vertical and horizontal axes, which are then averaged for multiple frames to reduce their uncertainty. c) A calculation of the change in the positions of the dark and bright ions as the radial field, $E_r$, is varied. Orange (blue) lines are the axial (radial) position of the bright (solid) and dark (dashed) ions. Green, blue, and red insets show the position of the dark ion (black open circle) compared to the bright ions (red circles) for different values of the radial electric field.}
    \label{fig:dark ion exp}
\end{figure*}

To test the validity of our dark-ion EMM compensation method, we use the technique developed in Refs.~\cite{schneider2012influence,gloger2015ion, saito2021measurement} with a small but useful modification. In these works, EMM compensation was performed by monitoring the position of a single bright ion in the two radial directions while changing the amplitude of the oscillating RF fields. In these works, the position of the ion along the imaging axis was retrieved by automated scanning of the imaging focus~\cite{gloger2015ion, saito2021measurement} or by inferring it from the size of the de-focused point-spread function~\cite{schneider2012influence}. Here, we show that for traps with a significant ratio between the frequencies of the two radial modes, we can compensate EMM in the two radial directions without the position information along the imaging axis. The possibility to compensate EMM in 2D from a 1D position information stems from the non-linear ion's trajectory when changing the RF amplitude~\cite{gloger2015ion}. 

In this work, we consider only EMM generated by the ion's radial displacement from the trap's RF center. Typically, this is the leading EMM term to compensate, which can acquire the largest value. Other EMM origins exist, such as out-of-phase EMM and axial EMM~\cite{Meir2017exp}, which are typically lower in magnitude. The presented method is insensitive to these EMM sources since they don't induce a constant shift from the trap's RF center. 

\section{Experiment - dark ion} 
We use a linear-segmented RF trap similar to the one described in Ref.~\cite{Meir2017exp}.   
We load ions to the trap from a heated oven loaded with pure calcium (Ca) grains. We use a 1+1' resonance-enhanced-ionization scheme to load ions to the trap from the flux of neutral atoms emerging from the heated oven. By tuning the first ionization beam to a resonance in neutral Ca, we can selectively ionize different Ca isotopes~\cite{lucas2004isotope}. We load a three-ion crystal made of two $^{40}$Ca$^+$ ions (bright ions) and one $^{44}$Ca$^+$ ion (dark ion) in the center. The cooling and fluorescence lasers are tuned to the resonance of the $^{40}$Ca$^+$ ions, hence the ``bright'' and '`dark'' nomenclatures. 

We fix the axial frequency of a single bright ion in the trap to $\omega^b_{ax}/2\pi=\SI{260.4}{\kilo\hertz}$. At a radial frequency,
\begin{equation}\label{eq:phase_transition_bright}
\omega^{b}_{r,zz}\approx\omega^b_{ax}\sqrt{\frac{4}{5}\left(1+\frac{2m_d}{m_b}\right)},
\end{equation}
the three-ion crystal transitions from a linear to a zig-zag configuration (see dark-ion appendix). Here, $m_b$ ($m_d$) is the bright (dark) ion mass. Eq.~\ref{eq:phase_transition_bright} states the trap frequency measured by a single bright ion. We can relate this frequency to the center-of-mass (COM) radial-mode frequency of a bright-dark-bright (bdb) crystal,
\begin{equation}\label{eq:radial_b_to_bdb}
    \omega_{r,\text{COM}}^{bdb}\approx\omega_r^b\frac{2}{3}\left(1+ \frac{m_b}{2m_d}\right).
\end{equation}
Using Eqs. \ref{eq:phase_transition_bright} and \ref{eq:radial_b_to_bdb}, we get $\omega_{r,zz}^{bdb}/2\pi\approx\SI{404}{\kilo\hertz}$ for the transition frequency from linear to zig-zag configuration in terms of the crystal's COM mode. Note that while the COM radial mode has a finite value near the transition to zig-zag, the bending radial mode, where the two bright ions move out of phase with respect to the dark ion, approaches zero (mode softening):
\begin{equation}\label{eq:bend}
    \omega_{r,\text{bend}}^{bdb}\sim\sqrt{(\omega_r^b)^2-(\omega_{r,zz}^b)^2}.
\end{equation} 

A static electric field $E_r$ in the trap's center will displace the ion according to~\cite{Berkeland1998}
\begin{equation}\label{eq:ion_shift}
    d_r\approx \frac{QE_r}{m\omega_r^2},
\end{equation}
where $Q$ is the electric charge of the ions. In a linear ion trap, to a leading order, $\omega_{ax}^2\propto m^{-1}$ while  $\omega_r^2\propto m^{-2}$ such that displacements from the center are different for the dark and bright ions in the radial direction (see Fig.~\ref{fig:dark ion exp}c). In addition, the displacement scales as $d_r\propto \omega^{-2}$. For that, the bending mode, whose frequency approaches zero near the zig-zag transition, governs the ion's displacement. 

We control the static electric field amplitude and orientation in the radial plane of the trap by applying voltages ($V_{x'}$, $V_{y'}$) on two compensation electrodes (see Fig.~\ref{fig:dark ion exp}a). As we scan the total excess field through its minimum, the dark ion is moved from one side of the crystal to the other as the sign of the external-field flips (see Fig.~\ref{fig:dark ion exp}c). When the trap is perfectly compensated, the dark ion is found directly between the two bright ions, pushing them to their maximum separation from each other. In the zig-zag configuration, however, the linear arrangement of ions is no longer stable, and as such, the dark ion will jump ``over'' the trap center. This jump occurs randomly when a fluctuation in the electric fields or a collision with background gasses grants it enough energy to bypass the potential barrier between the two sides of the crystal.

The radial plane of the trap supports two orthogonal modes ($\omega_h,\omega_l$), which are typically non-degenerate, $\omega_h>\omega_l$. We vary the radial frequencies' amplitudes and their orientation by changing the amplitude of the RF oscillating fields and the bias voltage (see Fig.~\ref{fig:dark ion exp}a). A suitable linear combination of the two compensation voltages allows us to push the ion toward each radial mode. 

We record the position of the ions on an EM-CCD camera (Fig.~\ref{fig:dark ion exp}b). The imaging system comprises an objective with a working distance of $\sim$30 \si{\mm} and a focal length of $\sim$37 \si{\mm}. The imaging-system magnification is measured to be $\sim$1.13 \si{\micro\metre}/pixel. The exposure time is 1 \si{\sec}. The image is projected on the trap's axial axis and on some projection of both the radial modes. In this experiment, we measure the axial distance between the two bright ions (see Fig.~\ref{fig:dark ion exp}b) from which we can infer the radial displacement of the dark ion from the axial axis. 

\section{Results - dark ion} 
\begin{figure}
    \centering
    \includegraphics[width=\linewidth]{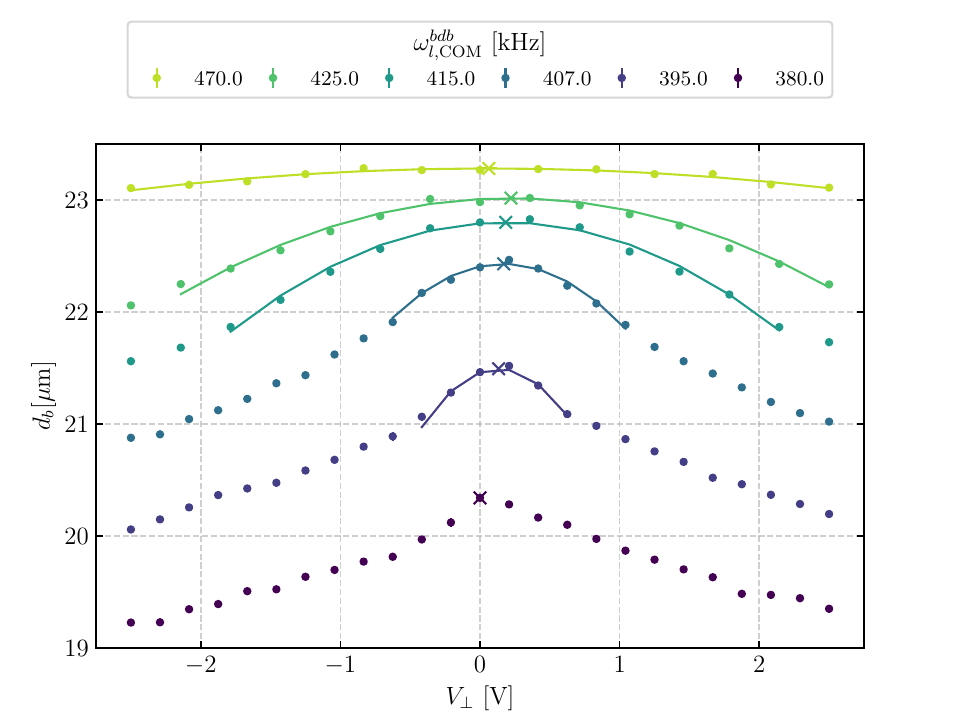}
    \caption{Bright ion separation, $d_b$, as function of an applied radial electric field for different low radial mode frequencies, $\omega_{l, \text{COM}}^{bdb}$ (legend). The electric field is applied along the direction of the low radial mode. The trap is compensated (not necessarily at the x-axis origin) when the distance between the bright ions is at its maximum. We extract the compensation value via a local parabola fit to the data (solid lines, cross indicate fit maximum). 
    The measurement peak becomes more pronounced, and the range of quadratic dependence narrows as the radial frequency approaches the transition to zig-zag. 
    Below the transition frequency to zig-zag ($\sim\SI{404}{\kilo\hertz}$), bright ions display a noticeable discontinuity in their positions as the crystal jumps between the zig and zag configurations, with the bright ion distance no longer fitting a parabola when looking far below the transition frequency.}
    \label{fig:sensitivity}
\end{figure}

\begin{figure*}
    \centering
    \includegraphics[trim={0cm 0cm 2.0cm 0cm},clip,width=\linewidth]{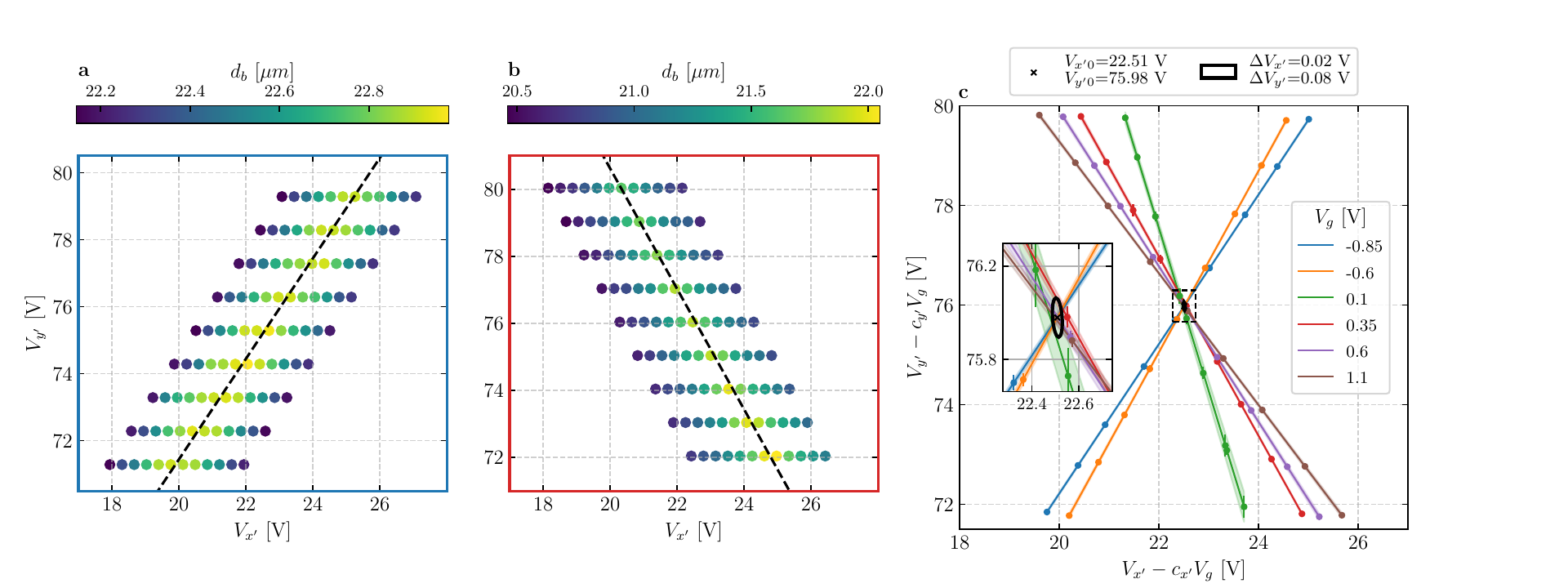}
    \caption{Dark ion EMM compensation. a-b) Bright ion distance (color scale) as a function of the compensation voltages ($V_{x'}$, $V_{y'}$) for different bias voltages.  Throughout the measurements, we kept the low-radial mode frequency at $\omega_{l,\text{COM}}^{bdb}=\SI{430}{\kilo\hertz}$ to avoid accidentally crossing the zig-zag transition frequency mid-measurement. The maximal ion distance follows a linear line in ($V_{x'}$, $V_{y'}$) space (dashed black line), the angle of which is determined by the radial-modes orientation. These lines are identical to those in (c) for bias voltages, $V_g=\SI{-0.85}{\V}$ (blue) and $V_g=\SI{0.35}{\V}$ (red), as denoted by the color of the frames. c) Compensation lines for different bias voltages (see legend). Each line is a linear fit as in (a-b), the points being the peaks of the ion distances found for each $V_{y'}$ in the 2D scans. The lines are shifted by the dependence of the compensation point on $V_g$ (Eq.~\ref{eq:Vg_mis}), so they all intersect at a single point ($c_{x'}=-0.23\pm0.07$, $c_{y'}=0.57\pm0.04$). The ellipse shows the one-sigma error in the estimation of $(V_{x'0},V_{y'0})$.}
    \label{fig:dark_ion_compensation}
\end{figure*}

The dependence of the distance between the bright ions on the EMM and the low radial-mode frequency is shown in Fig.~\ref{fig:sensitivity}. We vary the compensation voltages, $V_\perp=f(V_{x'}, V_{y'})$, to create a field along the direction of the low-frequency mode, $\omega_l$, and record the bright-ions positions. Here, $f$ is a linear function of the compensation-electrodes voltage, which we found experimentally (see discussion related to Fig.~\ref{fig:dark_ion_compensation}a). We repeat this procedure for different radial-mode trapping frequencies. The value of the crystal COM mode, $\omega_{l,\text{COM}}^{bdb}$, was found using the ``tickle'' method~\cite{Drewsen2004}.

As $\omega_{l,\text{COM}}^{bdb}$ approaches the zig-zag transition frequency ($\sim\SI{404}{\kilo\hertz}$, see Eq.~\ref{eq:radial_b_to_bdb}), the bending mode confinement approaches zero, making the radial position of the dark ion, and thus the axial size of the crystal, more sensitive to uncompensated EMM. 
This increase in sensitivity can be seen in the steepness of the peaks in the different curves in Fig.~\ref{fig:sensitivity}a: Far from the zig-zag transition, the curves look almost entirely flat, making the peak indiscernible. While moving closer to the critical point, the curves grow progressively steeper until the zig-zag transition point, where the measurement is most sensitive. 

Below the transition frequency to zig-zag configuration, the dark ion no longer passes through the center of the crystal when $V_\perp=0$, as this is not a stable configuration for a zig-zag chain. Instead, the crystal will randomly jump between the zig and zag configurations due to the potential barrier between them, making the estimation of the peak inherently unreliable. 


For traps with considerable ratio between the high and low radial frequency modes~\cite{Saito2024Information} as our own, the sufficient sensitivity for dark ion displacement is achieved in only one radial direction: while $\omega_l$ approaches the zig-zag transition frequency, $\omega_h$ remains far above it, such that the sensitivity of our measurement is diminished (in our trap, $\omega_{h,\text{COM}}^{bdb}\approx\SI{590}{\kilo\hertz}$ at the zig-zag transition). As a result, conducting a 2D scan of the bright ion distance vs. the compensation voltages results in a ``compensation line'', $V_{y'}=mV_{x'}+b$, for which the distance is maximized (see Fig.~\ref{fig:dark_ion_compensation}a-b). Following this line of minimal sensitivity, the ion is pushed towards the high radial mode, $\omega_h$. We cannot determine the location of the compensation point along the line from a single 2D scan.

To overcome this obstacle, we exploit the ability to tune the orientation of the radial modes in our trap~\cite{Saito2024Information}. By changing the bias voltage $V_g$, we can change the trap's mode orientation by $\sim70^\circ$ (see Fig.~\ref{fig:dark_ion_compensation}b). This rotation of $\omega_h$ and $\omega_l$ allows us to find new compensation lines,
\begin{equation}\label{eq:linear}
    V_{y'} = m_gV_{x'} + b_g,
\end{equation}
where $m_g$ and $b_g$ are the linear coefficients for a specific bias voltage, $V_g$.
The intersection of all these lines should give the value of ($V_{x'0}$, $V_{y'0}$) for which EMM is compensated.  

\begin{figure*}
    \centering
    \includegraphics[trim={4cm 4.0cm 4cm 3.0cm},clip, width=\linewidth]{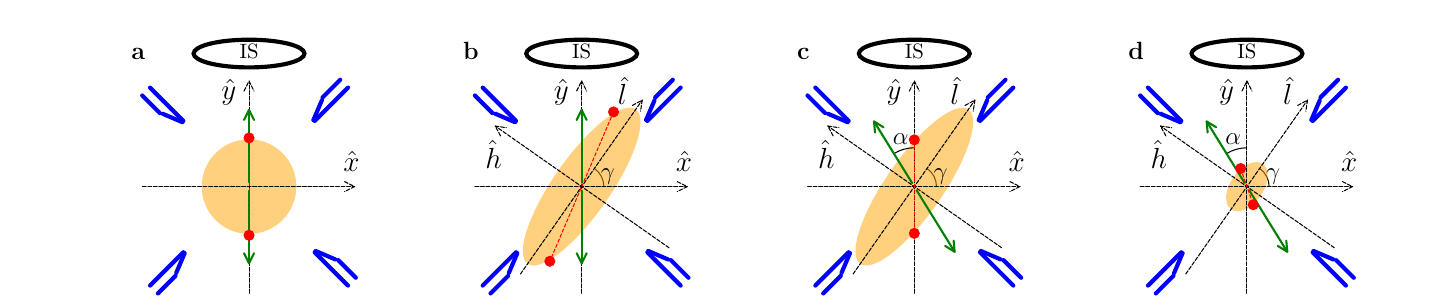}
    \caption{Single-ion radial displacement due to a radial stray electric field. 
    The imaging system (IS) is oriented along the y-axis and can detect only shifts in ion position along the x-axis.
    The radial mode frequencies ($\omega_l$, $\omega_h$) strength, ratio, and orientation ($\gamma$) are given by the orange ellipse. a) $\omega_l/\omega_h\approx1$. b-c) $\omega_l/\omega_h\approx 0.55$. d) $\omega_l/\omega_h\approx 0.77$. a-d) $\gamma= 55^\circ$.
    The stray electric field (green arrows) orientation is denoted by $\alpha$. a-b) $\alpha=0^\circ$. c-d) $\alpha\approx 32^\circ$. The two red dots connected by a dashed red line denotes the position of the ion for varying magnitude of the electric field.
    Blue blades illustrate the position and orientation of the trap electrodes. Through adjustment of the radial trapping frequencies, unobservable x-displacement (a,c) can be measured (b,d) - see text.}
    \label{fig:single_ion_exp}
\end{figure*}

However, by changing the bias voltage, we also create stray electric fields due to the misalignment of the bias and RF quadrupoles in the trap. This results in a linear dependence of the compensation point on the value of $V_g$:
\begin{equation}\label{eq:Vg_mis}
\begin{split}
    V_{x'g} = c_{x'}V_g+V_{x'0},\\
    V_{y'g} = c_{y'}V_g+V_{y'0}.
\end{split}
\end{equation}
Here, ($V_{x'g}$, $V_{y'g}$) are the compensation values for some value of $V_g$, ($V_{x'0}$, $V_{y'0}$) are the compensation values for $V_g=0$, and ($c_{x'}$, $c_{y'}$) are the linear coefficients, all of which are not known a-priori. 

Using Eq.~\ref{eq:Vg_mis} to ``shift'' all compensation lines (Eq.~\ref{eq:linear}) due to the effect of $V_g$, we get the following set of linear equations:
\begin{equation}\label{eq:linear_set}
    c_{y'}V_g+V_{y'0}=m_g(c_{x'}V_g+V_{x'0})+b_g.
\end{equation}
This set of equations has four ``free'' parameters: $c_{x'}$, $c_{y'}$, $V_{x'0}$, and $V_{y'0}$. Hence, by scanning the 2D compensation voltages for at least four bias voltages, we can extract the compensation points for any value of $V_g$. The results of this procedure with six different bias voltages are shown in Fig.~\ref{fig:dark_ion_compensation}c.

\section{Experiment - single ion}
To corroborate the EMM compensation results using a dark ion, we measure EMM by monitoring the position of a single bright ion on the camera for different amplitudes of the trapping RF fields~\cite{Berkeland1998,schneider2012influence,gloger2015ion, saito2021measurement}.
Here, we show that for 2D radial EMM compensation, only information on ion displacement along one radial direction is necessary. In our setup, this corresponds to the x-axis, which is perpendicular to the imaging axis. Our technique relies on two necessary conditions: 1) Non-degeneracy of the trap's radial modes. 2) The orientation of any of the radial modes should not align with the imaging optical axis. Both of these conditions are met in a typical ion trap, with the fulfillment of the second condition being ensured through a simple adjustment of $V_g$.

As can be seen from Eq.~\ref{eq:ion_shift}, the magnitude of the ion's shift from the trap center due to a stray electric field depends on the radial trapping frequency. Changing the amplitude of the radial trapping fields from maximum to minimum values (denoted as ``high'' and ``low'' RF amplitudes) while monitoring the ion's position shift is an efficient way to detect EMM~\cite{Berkeland1998, schneider2012influence,gloger2015ion, saito2021measurement}. 

In a symmetric trap, where the two radial modes are degenerate ($\omega_l=\omega_h)$, the ion is displaced along the direction of the stray electric field (see Fig.~\ref{fig:single_ion_exp}a). In case the field is directed along the imaging-system axis, it will be hard to detect a shift in the position of the ion. In a non-symmetric trap, where the two radial modes are non-degenerate ($\omega_l<\omega_h$), the ion shifts more towards the low radial mode (Fig.~\ref{fig:single_ion_exp}b). In this case, we can detect a shift in the ion position perpendicular to the imaging axis even though the electric field points along the imaging system axis. 

However, even for the non-symmetric trap case, an electric field oriented by an angle (see single-ion appendix),
\begin{equation}\label{eq:alpha0}
    \alpha_0 = \gamma - \arctan\left(\tan(\gamma)(\omega_l/\omega_h)^2\right),
\end{equation}
will lead to a non-detectable ion displacement along the imaging axis (Fig.~\ref{fig:single_ion_exp}c). Here, $\gamma$ is the orientation angle of the radial modes with respect to the lab frame (see Fig.~\ref{fig:single_ion_exp}b-d). Nevertheless, since the angle $\alpha_0$ depends on the radial-mode frequency ratio, $\omega_l/\omega_h$, changing this ratio (see single-ion appendix) allows us to induce detectable ion displacement along the x-axis for any stray-electric-field orientation  (Fig.~\ref{fig:single_ion_exp}d).

\begin{figure*}
    \centering
    \includegraphics[trim={0.0cm 0.3cm 0.0cm 5.8cm},clip, width=\linewidth] {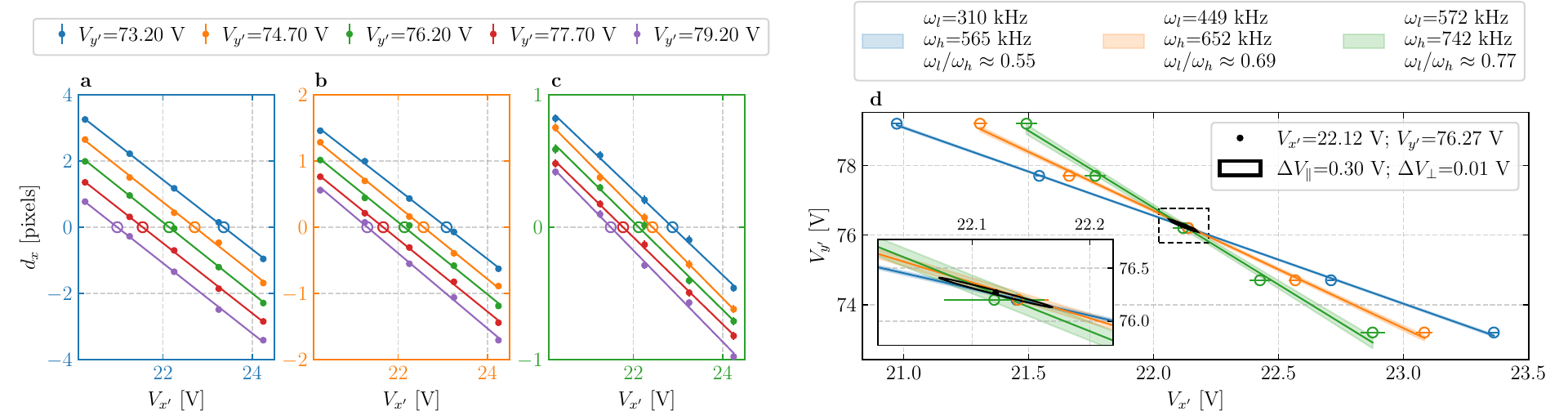}
    \caption{Single-ion EMM compensation. a-c) Dots are ion relative displacements between high to low RF amplitudes as a function of the $V_{x'}$ (x-axis) and $V_{y'}$ (color, left-side legend) compensation voltages. High RF amplitude corresponds to $\sim$\SI{1.5}{MHz} radial frequency. The three different low-RF-amplitude radial frequencies and ratios are given in the right-side legend. Errors are not visible at this scale, lines are linear fits, and open circles are estimates from fits for the zero-displacement $V_{x'}$ voltages.  d) Zero-displacement trajectories in the ($V_{x'}$,$V_{y'}$) space. Each color corresponds to a different low-RF amplitude (caption (a-c) box color). Open circles are the fit estimates extracted from (a-c) with the estimation error shown as horizontal lines. Lines are linear fits.  All lines cross at the EMM compensation point (black). The ellipse shows the 1$\sigma$ error in the compensation point estimation. The inset zooms into the crossing point. In this experiment, the bias voltage was fixed at a value of  $V_g=\SI{0.6}{V}$.}
    \label{fig:single_ion_results}
\end{figure*}

\section{Results - single ion} 
The results of EMM compensation with a single bright ion are presented in Fig.~\ref{fig:single_ion_results}. We scan the compensation-electrodes voltages, $V_{x'}$, $V_{y'}$, and record the ion's position for one high and three low RF amplitudes. The high RF amplitude corresponds to $\omega_l=1440$ \si{\kilo\hertz} and $\omega_h=1500$ \si{\kilo\hertz} for the low and high radial modes, respectively. The three low RF amplitudes correspond to $\omega_l=572, 449, 310$ \si{\kilo\hertz} for the low radial mode and $\omega_h=742, 652, 565$ \si{\kilo\hertz} for the high radial mode. We calculate the ion displacement along the x-axis, $d_x$, between the high RF amplitude and all low RF amplitudes (data points in Fig.~\ref{fig:single_ion_results}a-c). For each $V_{y'}$ compensation value, there exists a compensation value, $V_{x'}$, for which the ion displacement is zero (open circles in Fig.~\ref{fig:single_ion_results}a-c). 

We plot the zero-displacement compensation values in the ($V_{x'}$, $V_{y'}$) space (open circles in Fig.~\ref{fig:single_ion_results}d) and fit them to a linear line. We note that for each low RF amplitude (different radial-mode frequency ratios), the zero-displacement line follows a different angle in the ($V_{x'}$, $V_{y'}$) space. We find the zero-displacement lines intersection point (black dot in Fig.~\ref{fig:single_ion_results}d) and uncertainty (black ellipse in Fig.~\ref{fig:single_ion_results}d). The intersection point corresponds to the EMM compensated point where the radial stray electric field is minimized.

\section{Discussion} 
To compare our compensation scheme with other works, we relate the change in the compensation-electrodes voltage to the resulting electric field at the trap center. A typical uncertainty of $\Delta V_{x'/y'}\approx\SI{0.1}{V}$ in the compensation voltage corresponds to a stray electric-field uncertainty of $\Delta E_r\approx0.2\,\unit{\volt\per\meter}$ (see single-ion appendix). This level of compensation, which can be reached on a time scale of a few minutes, 
is consistent with the uncompensated stray-field magnitudes reported in typical ion-trap experiments.
A more qualitative comparison is difficult to perform due to the different trapping parameters used in various experiments. 

\begin{figure}
    \centering
    \includegraphics[width=\linewidth]{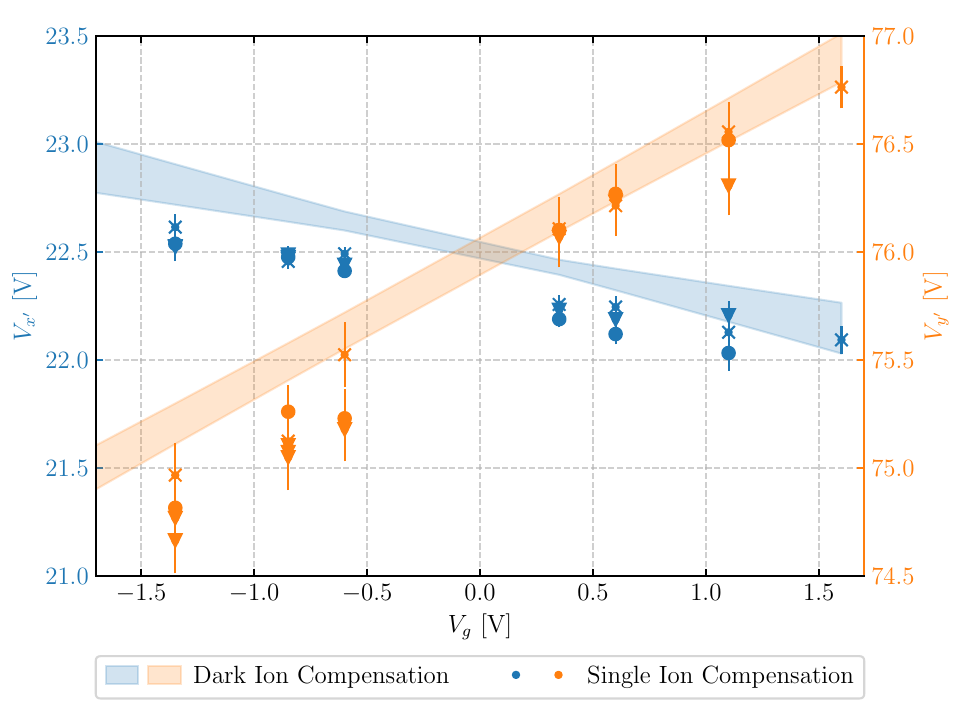}
    \caption{Comparison between the single ion (markers) and dark ion (shaded areas) compensation methods. The estimated $V_{x'}$ (blue, left y-axis) and $V_{y'}$ (orange, right y-axis) compensation voltages are shown as a function of the bias voltage $V_g$. Markers symbols corresponds to single-ion experiments performed before (cross) and after (circle, triangle) the dark-ion experiment.}
    \label{fig:comparison}
\end{figure}

A quantitative and direct comparison between the two compensation methods presented in this paper is shown in Fig.~\ref{fig:comparison}. While the two methods yield similar compensation uncertainties, there is a small discrepancy between them, which cannot be attributed to random error. As the results of the single-ion compensation are consistent before and after the dark-ion measurement, we can rule out drifts as the cause of the discrepancy. 

One possible systematic effect that differentiates between the two schemes is the presence of the scattering force from the fluorescence laser during the measurements. 
The scattering force acts as an additional effective stray field in the bright-ion scheme (on saturation, the maximal scattering field is $E_{{sc},\text{max}}\approx0.7\,\unit{\volt\per\meter}$). To compensate for the effect of the scattering force, we need to apply an equal and antiparallel field, $E_r=-E_{sc}$. In contrast, in the dark-ion scheme, the scattering force acts only on the bright ions, where the dark ion is unaffected by the scattering force, causing additional deformation in the crystal. To compensate for this deformation, we need to apply a compensation field $E_r\approx E_{sc}/(m_d/m_b-1)$ (see scattering-force appendix). This amounts to $E_r\approx10E_{sc}$ for our bright and dark ion masses. From the above discussion and the comparison made in Fig. \ref{fig:comparison}, we can give an upper bound to the scattering field in our experiment to be $E_{sc}\lesssim0.1\,\unit{\volt\per\meter}$.

In the bright-ion scheme, we perform EMM compensation independently for each bias voltage by exploiting the change in the low-to-high radial frequency ratio when increasing the RF amplitude (see legend in Fig. \ref{fig:single_ion_results}d).
In contrast, we need to perform several measurements for different bias voltages in the dark-ion scheme, resulting in a linear fit for the EMM compensation as a function of the bias voltage. 
This is because of the sensitivity scaling in the dark-ion scheme.
Since we are detecting axial contraction due to a field that shifts the ions radially, the sensitivity of the dark-ion scheme scales as $\frac{1}{E_r}\frac{d(d_b)}{dE_r}\propto (\omega_{r,\text{bend}}^{bdb})^{-4}$. 
Hence, we need to work close to the zig-zag transition to obtain sensitivity as the radial bending mode approaches zero frequency near the transition (Eq. \ref{eq:bend}). Our trap has a large low-to-high radial-frequency ratio ($\omega_l/\omega_h\approx0.7$). With this large ratio, we cannot obtain sensitivity to the high radial mode while working above the zig-zag transition in the low radial mode.
For more symmetric traps ($\omega_l\sim\omega_h$), it should be possible to simultaneously compensate EMM in the two radial directions in a single scan at a specific bias voltage. In contrast, the sensitivity of the single-ion method should decrease in more symmetric traps due to a smaller change in the low-to-high radial frequency ratio when changing the RF amplitude. 
For that, we expect the dark-ion method to be more beneficial in such scenarios. Unfortunately, we cannot meet these symmetric conditions in our experiment.

\section{Summary}
To summarize, we presented two EMM compensation methods that require only a fixed imaging system and continuous fluorescence detection, which makes them compatible and accessible to most ion-trapping experiments as is.  
Both methods overcome the difficulty of measuring the EMM in the direction parallel to the imaging system. The dark-ion method detects axial deformation of a bright-dark-bright ion crystal due to radial stray electric fields. In contrast, the bright-ion method exploits the curved ion trajectory due to changing trap asymmetry.
Both methods allow for a fair EMM compensation well below $\sim 1\,\unit{\volt\per\meter}$ of stray-electric-field amplitude.    

\section{Acknowledgments}
We thank Roee Ozeri for fruitful discussions and reading of this manuscript. We acknowledge the support of the Diane and Guilford Glazer Foundation Impact Grant for New Scientists, the research grant from the Donald Gordon Foundation and the Ike, Molly and Steven Elias Foundation, the research grant from the Laurie Kayden Foundation, the Center for New Scientists at the Weizmann Institute of Science, the Edith and Nathan Goldenberg Career Development Chair, the Israel Science Foundation (Grant No. 1010/22 and 1364/24), and the Minerva Stiftung with funding from the Federal German Ministry for Education and Research.

\input{output.bbl} 

\newpage
\section{Appendix - dark ion}
\begin{figure}
    \centering
    \includegraphics[width=\linewidth]{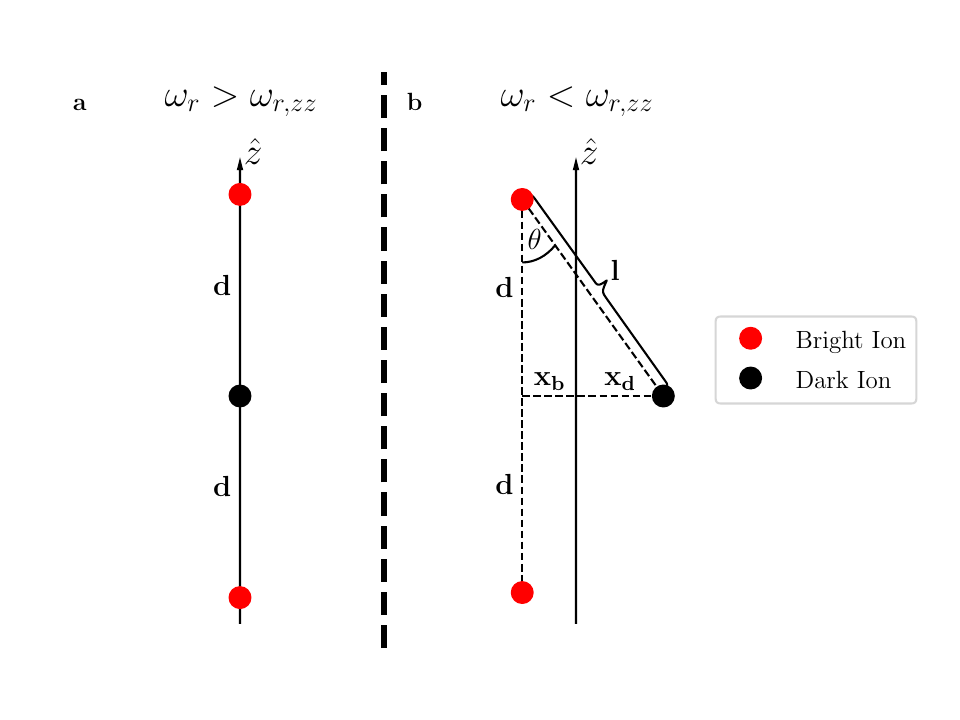}
    \caption{An ion crystal with two bright ions (red circles) and one dark ion (black circle) in the middle. a) Above the zig-zag transition frequency, the ions align in a linear configuration ($x_{b,d}=0$) at equilibrium. b) Below the zig-zag transition frequency, the equilibrium position of the ions deviates from a straight line ($x_{b,d}>0$).}
    \label{fig:appendix sketch}
\end{figure}

To derive the transition frequency from linear to zig-zag configuration (Eq.~\ref{eq:phase_transition_bright}), let us examine a three-ion crystal in a bright-dark-bright configuration with no external electric field (see Fig.~\ref{fig:appendix sketch}). 
In equilibrium, the forces acting on the bright ions in the axial and radial directions are 
\begin{align}
    &m_b(\omega^b_{ax})^2d=\frac{k_e}{(2d)^2}+\frac{k_e \cos(\theta)}{l^2}, \label{eq:bright ion axial} \\ 
    &m_b(\omega_{r}^b)^2x_b = \frac{k_e \sin(\theta)}{l^2}, \label{eq:bright ion radial}
\end{align}
while the radial forces on the dark ion are equal to
\begin{equation}\label{eq:dark ion radial}
    m_d(\omega_{r}^d)^2x_d=2\frac{k_e \sin(\theta)}{l^2}.
\end{equation}
Here, $d=d_b/2$ is the axial distance of the bright ions from the trap center. The radial distance of the bright (dark) ion from the trap center is $x_b$ ($x_d)$, where both are taken as positive numbers. From these definitions, the distance between the bright and dark ion, $l$, and their angle, $\theta$, follows, $d=l\cos(\theta)$ and $x_d+x_b=l\sin(\theta)$. The Coulomb constant, $k_e=Q^2/4\pi\varepsilon_0$, is defined together with the ions' charges for brevity.  

We assume the crystal is nearly linear, and therefore, the angle $\theta$ is small ($l\approx d$) such that
\begin{align}\label{eq:small angles}
    &\tan(\theta)\approx\sin(\theta)\approx \frac{x_d+x_b}{d}, \\
    &\cos(\theta)\approx1.
\end{align}
Under this approximation, from Eq.~\ref{eq:bright ion axial}, we get
\begin{equation}
m_b(\omega^b_{ax})^2d=\frac{5k_e}{4d^2}\rightarrow\frac{k_e}{d^3}=\frac{4}{5}m_b(\omega_{ax}^b)^2.
\end{equation}
Inserting this into the radial equations (Eqs.~\ref{eq:bright ion radial} and \ref{eq:dark ion radial}) yields
\begin{align}
&m_b(\omega_r^b)^2x_b=\frac{k_e(x_d+x_b)}{d^3}=\frac{4}{5}m_b(\omega_{ax}^b)^2(x_d+x_b),\\
&m_d(\omega_r^d)^2x_d=\frac{2k_e(x_d+x_b)}{d^3}=\frac{8}{5}m_b(\omega_{ax}^b)^2(x_d+x_b).
\end{align}

The system transforms to the zig-zag configuration upon the onset of a non-trivial solution ($x_d,x_b \neq 0$) to this set of equations, which gives rise to the following condition:
\begin{equation}\label{eq:critic}
\frac{4}{5}m_b(\omega_{ax}^b)^2=\frac{m_b(\omega_r^b)^2\cdot m_d(\omega_r^d)^2}{2m_b(\omega_r^b)^2+m_d(\omega_r^d)^2}.
\end{equation}
To further simplify the above expression, we neglect the DC and bias fields effect on the radial trapping frequencies such that
\begin{equation}\label{eq:app:simplify}
m_d(\omega_r^{d})^2=
\frac{m_b}{m_d}\cdot m_b(\omega_r^b)^2.
\end{equation}
Inserting this into Eq.~\ref{eq:critic}, we get the expression for critical frequency given in Eq. \ref{eq:phase_transition_bright} in the main text.

\section{Appendix - single ion}\label{app:single}
A single bright ion, mass $m$ and charge $Q$, is influenced by a radial stray electric field, $\mathbf{E}_r$, oriented by angle $\alpha$ from the imaging axis (y-axis). The ion's low and high radial trapping modes are denoted by $\omega_l$ and $\omega_h$ and are rotated by angle $\gamma$ from the lab's x-axis and y-axis respectively (see Fig.~\ref{fig:single_ion_exp}).

The displacements of the ion along each of its radial modes are given by,
\begin{align}
    &d_h = QE_r\cos(\gamma-\alpha)/m\omega_h^2 \\
    \nonumber 
    &d_l = QE_r\sin(\gamma-\alpha)/m\omega_l^2.
\end{align}
Transforming the above displacements to the lab frame yields,
\begin{align}\label{eq:dy}
    &d_y = \frac{QE_r}{m}\left[\frac{\sin(\gamma-\alpha)\sin\gamma}{\omega_l^2}+\frac{\cos(\gamma-\alpha)\cos\gamma}{\omega_h^2}\right] \\
    \label{eq:dx}
    &d_x = \frac{QE_r}{m}\left[\frac{\sin(\gamma-\alpha)\cos\gamma}{\omega_l^2}-\frac{\cos(\gamma-\alpha)\sin\gamma}{\omega_h^2}\right].
\end{align}
The condition for no displacement in the x-axis, $d_x=0$, determines the angle of undetected stray electric field,
\begin{equation}
    \alpha_0 = \gamma - \tan^{-1}\left(\tan(\gamma)(\omega_l/\omega_h)^2\right).
\end{equation}
Note that this angle depends on ratio of low and high radial-mode frequencies, $\omega_l/\omega_h$.

Due to the symmetric positions of the compensation electrodes (Fig. \ref{fig:dark ion exp}), we can assume that the electric field created by the compensation electrodes follows:
\begin{equation}
    E_r=\eta \sqrt{V_{x'}^2+V_{y'}^2}.
\end{equation}
Here, $\eta$ is the electric-field response to a change in the compensation voltage.

We can use Eq. \ref{eq:dx} to fit $d_x$ as a function of $V_{x'}$ and $V_{y'}$, thereby finding $\eta$ from the measurements shown in Fig. \ref{fig:single_ion_results}. Since we cannot directly observe $d_y$, there is a degeneracy between the amplitude of the response and the angles $\gamma$ and $\alpha$, necessitating the use of measurements with multiple $\omega_l/\omega_h$ ratios and the additional assumptions that the fields generated by $V_{x'}$ and $V_{y'}$ are indeed orthogonal and scale identically with the voltage. From this fit we estimate $\eta=2.2\pm0.1\,\unit{\per\meter}$

The radial-mode frequencies are determined by the trap parameters
\begin{equation}
    \omega_{l/h}=\frac{\Omega}{2}\sqrt{a_{l/h}+\frac{q_{l/h}^2}{2}}.
\end{equation}
Here, $\Omega/2\pi=\SI{14.5456}{MHz}$ is the RF drive frequency, $q_l$ and $a_l$ ($q_h$ and $a_h$) are the trap's RF and DC parameters for the low (high) radial mode, respectively. The trap's parameters are proportional to the applied RF and DC voltages. In our trap, the RF drive is symmetric in the radial plane, $q_l\approx q_h$, while the DC fields are not, $a_l\neq a_h$. For that, we have a considerable asymmetry in the radial-mode frequencies, $\omega_l<\omega_h$.  

Due to misalignment between the DCs and RF quadrupoles, EMM compensation is performed with a fixed DC voltage as to avoid additional changes in the electric field. To change the ratio of the two radial modes, we change the trap's $q$ parameter by changing the RF amplitude. Note that by doing so we also increase the absolute value of the radial modes, thus reducing the sensitivity for detecting ion displacements.  

\section{Appendix - scattering force}\label{app:scattering}
We estimate the maximal scattering force on a single bright ion by
\begin{equation}
    QE_{sc,\text{max}}=\hbar k A_{sp}/2.
\end{equation}
Here, $\hbar$ is the reduced Planck's constant, $k=2\pi/\lambda$ is the laser's k-vector with $\lambda$ the laser wavelength of 397 nm, and $A_{sp}\approx \qty{1.36e8}{\per\second}$~\cite{UDportal} is the transition rate between the 4S$_{1/2}$ and the 4P$_{1/2}$ states in $^{40}$Ca$^+$. 

In the dark-ion method, the scattering force only affects the bright ions; hence, it deforms the crystal. In this case of uneven forces on the dark and bright ions, we want to calculate the compensation field necessary to make the crystal linear again. In such a case ($x_b=x_d\equiv x$), the radial forces on the bright and dark ions are
\begin{align}
    &m_b(\omega_{r}^b)^2x = Q(E_r+E_{sc}), \label{eq:app:brE} \\ 
    &m_d(\omega_{r}^d)^2x=QE_r. \label{eq:app:drE}
\end{align}
Dividing the two above equations, we get, after some rearrangement,
\begin{equation}
    E_r=E_{sc}/ \left(\frac{m_b(\omega_{r}^b)^2}{m_d(\omega_{r}^d)^2}-1\right),
\end{equation}
which can be further simplified using Eq. \ref{eq:app:simplify} to
\begin{equation}
    E_r\approx E_{sc}/ \left({m_d}/{m_b}-1\right).
\end{equation}

\end{document}

%% file: output.bbl
%